# Non-Bernoulli Perturbation Distributions for Small Samples in Simultaneous Perturbation Stochastic Approximation

Xumeng Cao
xcao7@jhu.edu
Department of Applied Mathematics and Statistics
Johns Hopkins University, MD, USA

**ABSTRACT**

**Simultaneous perturbation stochastic approximation (SPSA) has proven to be efficient for recursive optimization. SPSA uses a centered difference approximation to the gradient based on two function evaluations regardless of the dimension of the problem. Typically, the Bernoulli ±1 distribution is used for perturbation vectors and theory has been established to prove the asymptotic optimality of this distribution. However, optimality of the Bernoulli distribution may not hold for small-sample stochastic approximation (SA) runs. In this paper, we investigate the performance of the segmented uniform as a perturbation distribution for small-sample SPSA. In particular, we conduct a theoretical analysis for one iteration of SA, which is a reasonable starting point and can be used as a basis for generalization to other small-sample SPSA settings with more than one iteration. In this work, we show that the Bernoulli distribution may not be the best choice for perturbation vectors under certain choices of parameters in small-sample SPSA.**

1. INTRODUCTON

Simultaneous perturbation stochastic approximation (SPSA) has proven to be an efficient stochastic approximation approach (see [6, 7 and 9]). It has wide applications in areas such as signal processing, system identification and parameter estimation (see www.jhuapl.edu/SPSA/ and [2, 8]). The merit of SPSA follows from the construction of the gradient approximation, where only two function evaluations are needed for each step of the gradient approximation regardless of the dimension of the unknown parameter. As a result, SPSA reduces computation demand as compared to the finite difference (FD) method, which requires $2p$ function evaluations to achieve each step of the gradient approximation, where $p$ is the dimension of the problem (see [8], Chapters 6 and 7). Obviously, the savings in computation with SPSA is more significant as $p$ gets larger.

The implementation of SPSA involves perturbation vectors. Typically, the Bernoulli ±1 distribution is used for the components of the perturbation vectors. This distribution is easy to implement and has been proven asymptotically most efficient (see [5]). As a result, for large-sample SPSA, the Bernoulli distribution is the best choice for the perturbation vectors. However, one might be curious if this optimality remains when only small-sample stochastic approximation (SA) is allowed. Small-sample SA appears commonly in practice where it is expensive, either physically or computationally, to evaluate system performances. For example, it might be very costly to run experiments on a complicated control system. Under such circumstances, a limited number of function evaluations are available for SA. Unlike with large-sample SPSA, one might not be confident that the Bernoulli distribution is the best choice for the perturbation vectors in small-sample SPSA.

In this note, we discuss the effective perturbation distributions for SPSA with limited samples. Specifically, we consider the segmented uniform (SU) distribution as a representative of non-Bernoulli distributions. The SU distribution has nice properties of easy manipulation both analytically and numerically. For instance, it has both a density function and a distribution function in closed form, making analytical computations possible. Moreover, it does not take much effort to generate SU random variables due to the nature of the SU density, resulting in time-efficient numerical analysis. In our discussion, we focus on one-iteration SPSA, which is a special case of small-sample SPSA. As a finite-sample analogue to asymptotic cases, the one-iteration case is a good starting point as it is easier to analyze and still captures insightful properties of general small-sample SPSA. Through the analysis of the one-iteration case, we get insights on the behavior of other small samples in the hope that the analysis can be generalized to more than one iteration case. In fact, we demonstrate numerically that the one-iteration theoretical conclusions do apply to more than one iteration.

Discussion and research on non-Bernoulli perturbation distributions in SPSA have been found in the literature (see [1, 3]). In [1], numerical experiments along with rigorous convergence proofs indicate that deterministic perturbation sequences show promise for significantly faster convergence under certain circumstances; while in [3], conjecture is made based on empirical results that the Bernoulli distribution maintains optimality for small-sample analysis given an optimal choice of parameters. However, no theoretical foundation is provided to validate this conjecture. The application of non-Bernoulli perturbations in SPSA is discussed in [4] and [8, Section 7.3].

## 2. METHODOLOGY

### 2.1 Problem Formulation

Let $\boldsymbol{\theta} \in \Theta \subseteq R^p$ denote a vector-valued parameter of interest, where $\Theta$ is the parameter space and $p$ is the dimension of $\boldsymbol{\theta}$. Let $L(\boldsymbol{\theta})$ be the loss function, which is observed in the presence of noise: $y(\boldsymbol{\theta}) = L(\boldsymbol{\theta}) + \varepsilon$, where $\varepsilon$ is i.i.d noise, with mean zero and variance $\sigma^2$, $y(\boldsymbol{\theta})$ is the observation of $L(\boldsymbol{\theta})$ with noise $\varepsilon$. The problem is to

$$\min_{\theta \in \Theta} L(\theta). \tag{1}$$

The stochastic optimization algorithm to solve (1) is given by the following iterative scheme:

$$\hat{\theta}_{k+1} = \hat{\theta}_k - a_k \hat{g}_k(\hat{\theta}_k), \tag{2}$$

where $\hat{\theta}_k$ is the estimate of $\theta$ at iteration $k$ and $\hat{g}_k(\cdot) \in R^p$ represents an estimate of the gradient of $L$ at iteration $k$. The scalar-valued step-size sequence $\{a_k\}$ is nonnegative, decreasing, and converging to zero. The generic iterative form of (2) is analogous to the steepest descent algorithm for deterministic problems.

## 2.2 Perturbation Distribution for SPSA

SPSA uses simultaneous perturbation to estimate the gradient of $L$. The efficiency of this method is that it requires only two function evaluations at each iteration, as compared to $2p$ for the FD method (see [8], Chapters 6 and 7). Let $\Delta_k$ be a vector of $p$ scalar-valued independent random variables at iteration $k$:

$$\Delta_k = [\Delta_{k1}, \Delta_{k2}, ..., \Delta_{kp}]^T.$$

Let $c_k$ be a sequence of positive scalars. The standard simultaneous perturbation form for the gradient estimate is as follows:

$$\hat{g}_k(\hat{\theta}_k) = \begin{bmatrix} \dfrac{y(\hat{\theta}_k + c_k \Delta_k) - y(\hat{\theta}_k - c_k \Delta_k)}{2 c_k \Delta_{k1}} \\ \vdots \\ \dfrac{y(\hat{\theta}_k + c_k \Delta_k) - y(\hat{\theta}_k - c_k \Delta_k)}{2 c_k \Delta_{kp}} \end{bmatrix}. \tag{3}$$

To guarantee the convergence of the algorithm, certain assumptions on $\Delta_k$ should be satisfied:

**A1.** $\{\Delta_{ki}\}$ are independent for all $k$, $i$, and identically distributed for all $i$ at each $k$.

**A2.** $\{\Delta_{ki}\}$ are symmetrically distributed about zero and uniformly bounded in magnitude for all $k$, $i$.

**A3.** $E\left[\left(y(\hat{\theta}_k \pm c_k \Delta_k)/\Delta_{ki}\right)^2\right]$ is uniformly bounded over $k$ and $i$.

Condition A3 has an important relationship with the finite inverse moments of the elements of $\Delta_k$ (see [8], p. 184). An important part of SPSA is the bounded inverse moments condition for the $\Delta_{ki}$. Valid distributions include the Bernoulli ±1, the segmented uniform, the U-shape distribution and many others (see [8], p. 185). Two common mean-zero distributions that do not satisfy the bounded inverse moments condition are the symmetric uniform and the mean-zero normal distributions. The failure of both these distributions is a consequence of the amount of probability mass near zero.

In the discussion that follows, we compare the segmented uniform (SU) distribution with the Bernoulli ±1 distribution. To guarantee that the two distributions have the same mean and variance, the domain of SU, following from basic statistics and simple algebra, is given as $\left(-(19+3\sqrt{13})/20, -(19-3\sqrt{13})/20\right) \cup \left((19-3\sqrt{13})/20, (19+3\sqrt{13})/20\right)$, which is approximately $(-1.4908, -0.4092) \cup (0.4092, 1.4908)$, see Figure 1. In our analysis, the sequences $\{a_k\}$ and $\{c_k\}$ take standard forms: $a_k = a/(k+2)^{0.602}$, $c_k = c/(k+1)^{0.101}$, where $a$ and $c$ are predetermined constants.

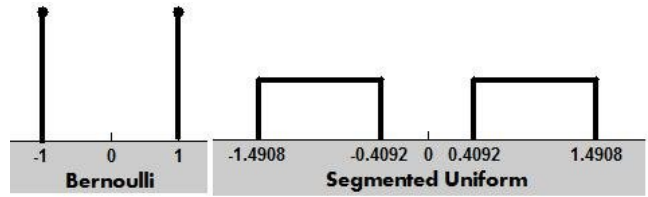

**Figure 1**: Mass/ probability density functions of the Bernoulli ±1 and the segmented uniform distributions. Both distributions have mean 0 and variance 1.

Moments of perturbations under two distributions are summarized below in Table 1. These moments will be used in Section 3. Subscripts $i$ and $j$ denote the elements of $\Delta_0$ and $i \neq j$. The derivation follows from basic statistics and simple algebra.

Table 1: Moments of perturbations under two distributions

| Expectation | Bernoulli | SU |
|---|---|---|
| $E(\Delta_{0i})$ | 0 | 0 |
| $E(\Delta_{0i}/\Delta_{0j})$ | 0 | 0 |
| $E(\Delta_{0i}^2/\Delta_{0j}^2)$ | 1 | 100/61 |
| $E(1/\Delta_{0i}^2)$ | 1 | 100/61 |

## 3. THEORETICAL ANALYSIS

In this section, we provide conditions under which SU outperforms the Bernoulli distribution. To specifically analyze the development of the algorithm, we consider the extreme example of small samples where only one iteration takes place in SPSA, that is, $k = 1$. We start with this simple case as a basis for possible generalization for larger values of $k$, where the analysis is more complicated. In our analysis, mean squared error (MSE) is used to compare the performance of two distributions.

Before we present the results, let us define necessary notations. Subscripts $S$, $B$ denote SU and the Bernoulli distribution, respectively, e.g. $a_{0S}$ denotes the value of $a_0$ under SU distribution; $L_i$ is the first derivatives of $L$ with respect to the $i$th component of $\boldsymbol{\theta}$, all first derivatives are evaluated at the starting point $\hat{\boldsymbol{\theta}}_0$; $\hat{\theta}_{0i}$ and $\theta_i^*$ are the $i$th component of $\hat{\boldsymbol{\theta}}_0$ and $\boldsymbol{\theta}^*$, respectively, where $\boldsymbol{\theta}^*$ is the true value of $\boldsymbol{\theta}$. Following the theorem statement below, we provide some interpretation of the main condition.

**Theorem 1**

Consider loss function $L(\boldsymbol{\theta})$ with continuous third derivatives. For one iteration of SPSA, the SU distribution produces a smaller MSE between $\hat{\boldsymbol{\theta}}_1$ and $\boldsymbol{\theta}^*$ than the Bernoulli $\pm 1$ distribution if the starting point and the relevant coefficients $(a_0, c_0, \sigma^2)$ are such that the following is true:

$$\left[\left(\frac{100}{61}p - \frac{39}{61}\right)a_{0S}^2 - pa_{0B}^2\right]\sum_{i=1}^{p}L_i^2$$
$$+ (a_{0S} - a_{0B})\left[\frac{p\sigma^2}{2c_{0B}^2}(a_{0S} + a_{0B}) - 2\sum_{i=1}^{p}(\hat{\theta}_{0i} - \theta_i^*)L_i\right]$$
$$- pa_{0S}^2\sigma^2\left(\frac{1}{2c_{0B}^2} - \frac{50}{61c_{0S}^2}\right) + O(c_0^2) < 0, \tag{4}$$

where the $O(c_0^2)$ term is due to the higher order Taylor expansion.

**Remark 1:** The choice of the coefficients is not arbitrary. For example, $a_0$ and $c_0$ should be picked according to the standard tuning process (see [8], Section 7.5); the starting point should stay in a reasonable range given any prior information for the problem. To best use the result of Theorem 1, one should follow these standards rather than arbitrarily picking the coefficients to make (4) true.

**Remark 2:** If the gains $c_{0S}$ and $c_{0B}$ are small enough such that $O(c_0^2)$ is negligible, the following conditions ((a) and (b)) would be sufficient for (4) to hold:

(a) The ratios of the gain sequences have the following relations:

$$a_{0S}/a_{0B} < \sqrt{p/(100p/61 - 39/61)} < 1$$

$$c_{0B}/c_{0S} < \sqrt{61/100} \approx 0.781$$

(b) The function is relatively flat and the starting point is not too far away from the true minimum. In particular, the following inequality is true:

$$2\sum_{i=1}^{p}(\hat{\theta}_{0i} - \theta_i^*)L_i < \frac{p\sigma^2}{2c_{0B}^2}(a_{0S} + a_{0B}).$$

*Proof of Theorem 1*: By (2) and (3), the updated estimate of $\theta$ after one iteration is

$$\hat{\theta}_1 = \hat{\theta}_0 - a_0 \frac{y(\hat{\theta}_0 + c_0\Delta_0) - y(\hat{\theta}_0 - c_0\Delta_0)}{2c_0} \times \left[\Delta_{01}^{-1}, \ldots, \Delta_{0p}^{-1}\right]^T$$

$$= \hat{\theta}_0 - a_0 \frac{L(\hat{\theta}_0 + c_0\Delta_0) + \varepsilon^+ - (L(\hat{\theta}_0 - c_0\Delta_0) + \varepsilon^-)}{2c_0} \times \left[\Delta_{01}^{-1}, \ldots, \Delta_{0p}^{-1}\right]^T, \quad (5)$$

where $\varepsilon^+$ and $\varepsilon^-$ are the corresponding noise. By a Taylor expansion of the third order,

$$L(\hat{\theta}_0 + c_0\Delta_0) - L(\hat{\theta}_0 - c_0\Delta_0) = 2c_0 \sum_{i=1}^{p} L_i \Delta_{0i} + O(c_0^3), \quad (6)$$

where the $O(c_0^3)$ term is due to the higher order Taylor expansion. Specifically,

$$O(c_0^3) = \frac{1}{6} c_0^3 \sum_{i=1}^{p} \sum_{j=1}^{p} \sum_{k=1}^{p} \left[\left(L_{ijk}(\tilde{\theta}) + L_{ijk}(\tilde{\tilde{\theta}})\right) \Delta_{0i} \Delta_{0j} \Delta_{0k}\right], \quad (7)$$

where $L_{ijk}$ denotes the third derivatives of $L$ with respect to the elements $i, j, k$ of $\theta$, $\tilde{\theta}$ and $\tilde{\tilde{\theta}}$ are the intermediate points between $\hat{\theta}_0$ and $\hat{\theta}_0 + c_0\Delta_0$, $\hat{\theta}_0$ and $\hat{\theta}_0 - c_0\Delta_0$, respectively.

Given (5), (6) and (7), and following from algebraic calculation and necessary rearrangements, we compute the difference in MSE $E\left(\|\hat{\theta}_1 - \theta^*\|^2\right) = E\left(\left(\hat{\theta}_1 - \theta^*\right)^T \left(\hat{\theta}_1 - \theta^*\right)\right)$ under two distributions as follows:

$$E_S\left(\|\hat{\boldsymbol{\theta}}_1 - \boldsymbol{\theta}^*\|^2\right) - E_B\left(\|\hat{\boldsymbol{\theta}}_1 - \boldsymbol{\theta}^*\|^2\right)$$

$$= \left[\left(\frac{100}{61}p - \frac{39}{61}\right)a_{0S}^2 - pa_{0B}^2\right]\sum_{i=1}^{p} L_i^2$$

$$+ (a_{0S} - a_{0B})\left[\frac{p\sigma^2}{2c_{0B}^2}(a_{0S} + a_{0B}) - 2\sum_{i=1}^{p}(\hat{\theta}_{0i} - \theta_i^*)L_i\right]$$

$$- pa_{0S}^2\sigma^2\left(\frac{1}{2c_{0B}^2} - \frac{50}{61c_{0S}^2}\right) + O(c_0^2) \tag{8}$$

The derivation of (8) involves the computation of relevant moments, which are summarized in Table 1. □

Condition (8) in Theorem 1 may be hard to check for general problems due to the unknown analytical form of the higher order term $O(c_0^2)$. However, if we know more information about the loss function $L$, condition (8) can be replaced by a sufficient condition, which is easier to manipulate in practice.

**Corollary 1**

If we assume an upper bound for the magnitude of the third derivatives of $L$, say, $|L_{ijk}(\bullet)| \leq M$ for all $i, j, k$, where $M$ is a constant, we can establish an upper bound $U$ for the term $O(c_0^2)$ in (8), i.e. $O(c_0^2) \leq U$. As a result, a more conservative condition for the superiority of SU is

$$\left[\left(\frac{100}{61}p - \frac{39}{61}\right)a_{0S}^2 - pa_{0B}^2\right]\sum_{i=1}^{p} L_i^2$$

$$+ (a_{0S} - a_{0B})\left[\frac{p\sigma^2}{2c_{0B}^2}(a_{0S} + a_{0B}) - 2\sum_{i=1}^{p}(\hat{\theta}_{0i} - \theta_i^*)L_i\right]$$

$$- pa_{0S}^2\sigma^2\left(\frac{1}{2c_{0B}^2} - \frac{50}{61c_{0S}^2}\right) + U < 0, \tag{9}$$

where $U$ is defined as:

$$U = (4a_{0S}^2 c_{0S}^2 + a_{0B}^2 c_{0B}^2) M \left( \sum_{i=1}^{p} |\hat{\theta}_{0i} - \theta_i^*| \right) (p-1)^2 + \frac{1}{20} a_{0S}^2 c_{0S}^4 M^2 p^7 a_{0S}$$

$$+ \frac{1}{3}(a_{0S}^2 c_{0S}^3 + a_{0B}^2 c_{0B}^3) M p^5 \max_i L_i(\hat{\theta}_0). \tag{10}$$

*Proof*: Given (7) and the assumption that $|L_{ijk}(\bullet)| \leq M$ for all $i, j, k$, we derive an upper bound $U$ for the term $O(c_0^2)$ as in (10). To derive (10), we should first find the explicit form of the term $O(c_0^2)$ in (8) as follows:

$$O(c^2) = E_S \left[ a_{0S}^2 V_S^T V_S - 2a_{0S}(\hat{\theta}_0 - \theta^* - a_{0S} W_S)^T V_S \right] - E_B \left[ a_{0B}^2 V_B^T V_B - 2a_{0B}(\hat{\theta}_0 - \theta^* - a_{0B} W_B)^T V_B \right],$$

where for $h = S$ or $B$, as appropriate,

$$W_h = \frac{2c_{0h} \sum_{i=1}^{p} L_i \Delta_{0i} + \varepsilon^+ - \varepsilon^-}{2c_{0h}} \times \left[ \Delta_{01}^{-1}, ..., \Delta_{0p}^{-1} \right]^T;$$

$$V_h = \frac{1}{12} c_{0h}^2 \sum_{i=1}^{p} \sum_{j=1}^{p} \sum_{k=1}^{p} \left[ \left( L_{ijk}(\tilde{\theta}) + L_{ijk}(\tilde{\tilde{\theta}}) \right) \Delta_{0i} \Delta_{0j} \Delta_{0k} \right] \times \left[ \Delta_{01}^{-1}, ..., \Delta_{0p}^{-1} \right]^T.$$

Given the upper bound in (10), it follows immediately that (9) is a sufficient and more conservative condition for the superiority (smaller MSE) of SU. □

Notice that if $L$ is quadratic, the higher order terms in (6) and (8) vanish, resulting in the following simpler form of the condition in Theorem 1.

**Corollary 2**

For a quadratic loss function $L$, the SU distribution produces a smaller MSE between $\hat{\theta}_1$ and $\theta^*$ than the Bernoulli ±1 distribution for one-iteration SPSA if the following holds:

$$\left[\left(\frac{100}{61}p - \frac{39}{61}\right)a_{0S}^2 - pa_{0B}^2\right]\sum_{i=1}^{p}L_i^2$$

$$+(a_{0S} - a_{0B})\left[\frac{p\sigma^2}{2c_{0B}^2}(a_{0S} + a_{0B}) - 2\sum_{i=1}^{p}(\hat{\theta}_{0i} - \theta_i^*)L_i\right]$$

$$-pa_{0S}^2\sigma^2\left(\frac{1}{2c_{0B}^2} - \frac{50}{61c_{0S}^2}\right) < 0.$$

If $p = 2$, the special form of Corollary 2 becomes the following, which we use in the numerical example 4.1 below:

**Corollary 3**

For a quadratic loss function with $p = 2$, the SU distribution produces a smaller MSE between $\hat{\theta}_1$ and $\theta^*$ than the Bernoulli $\pm 1$ distribution for one-iteration SPSA if the following holds:

$$\left[\frac{161}{61}a_{0S}^2 - 2a_{0B}^2\right]\left(L_1^2 + L_2^2\right)$$

$$+(a_{0S} - a_{0B})\left[\frac{\sigma^2}{c_{0B}^2}(a_{0S} + a_{0B}) - 2\left(L_1(\hat{\theta}_{01} - \theta_1^*) + L_2(\hat{\theta}_{02} - \theta_2^*)\right)\right]$$

$$-a_{0S}^2\sigma^2\left(\frac{1}{c_{0B}^2} - \frac{100}{61c_{0S}^2}\right) < 0. \tag{11}$$

## 4. NUMERICAL EXAMPLE

### 4.1 Quadratic loss function

Consider the quadratic loss function $L(\theta) = t_1^2 - t_1 t_2 + t_2^2$, where $\theta = [t_1, t_2]^T$, $\sigma^2 = 1$, $\hat{\theta}_0 = [0.3, 0.3]^T$, $a_S = 0.00167$, $a_B = 0.01897$, $c_S = c_B = 0.1$, i.e. $a_{0S} = a_S / (0+2)^{0.602} = 0.0011$, $a_{0B} = a_B / (0+2)^{0.602} = 0.01252$, $c_{0S} = c_S / (0+1)^{0.101} = 0.1$, $c_{0B} = c_B / (0+1)^{0.101} = 0.1$, i.e., the parameters are chosen according to the tuning process (see [8], Section 7.5). The left hand side of (11) is calculated as $-0.0114$, which satisfies the condition of Corollary 3, meaning SU outperforms the Bernoulli for $k = 1$. Now let us check this result with numerical simulation. We approximate the MSEs by averaging over $3\times 10^7$ independent sample squared errors. Results are summarized in Table 2.

**Table 2**: Results for quadratic loss functions

| Number of iterations | MSE for Bernoulli | MSE for SU | P-value |
|---|---|---|---|
| $k=1$ | 0.1913 | 0.1798 | $<10^{-10}$ |
| $k=5$ | 0.2094 | 0.1796 | $<10^{-10}$ |
| $k=10$ | 0.1890 | 0.1786 | $<10^{-10}$ |
| $k=1000$ | 0.0421 | 0.1403 | $>1-10^{-10}$ |

In Table 2, for each iteration count $k$, the MSEs $E\left(\left\|\hat{\boldsymbol{\theta}}_1 - \boldsymbol{\theta}^*\right\|^2\right)$ are approximated by averaging over $3\times 10^7$ independent sample squared errors. P-values are derived from standard matched-pairs $t$-tests for comparing two population means, which in this case are the MSEs for the Bernoulli and SU. For $k = 1$, the difference between MSEs under SU and the Bernoulli is −0.0115 (as compared to the theoretical value of −0.0114 computed from the expression in (11)), with the corresponding P-value being almost 0, which shows a strong indication that SU is preferred to the Bernoulli for $k = 1$.

We also notice that the advantage of SU holds for $k = 5$ and $k = 10$ in this example. In fact, the better performance of SU for $k > 1$ has been observed in other examples as well (e.g., [4] and [8, Exercise 7.7]). Thus, even though this paper only provides the theoretical foundation for $k = 1$ case, it might be possible to generalize the theory to $k > 1$ provided that $k$ is not too large a number.

### 4.2 Non-quadratic loss function

Consider the loss function $L(\boldsymbol{\theta}) = t_1^4 + t_1^2 + t_1 t_2 + t_2^2$, where $\boldsymbol{\theta} = [t_1, t_2]^T$, $\sigma^2 = 1$, $\hat{\boldsymbol{\theta}}_0 = [1,1]^T$, the tuning process (see [8], Section 7.5) results in $a_S = 0.05$, $a_B = 0.15$, $c_S = c_B = 1$. We estimate the MSEs by averaging over $10^6$ independent sample squared errors. Results are summarized in Table 3.

**Table 3**: Results for non-quadratic loss functions

| Number of iterations | MSE for Bernoulli | MSE for SU |
|---|---|---|
| k=1 | 1.7891 | 1.5255 |
| k=2 | 1.2811 | 1.2592 |
| k=5 | 0.6500 | 0.9122 |
| k=1000 | 0.0024 | 0.0049 |

In Table 3, for each iteration count $k$, the MSEs $E\left(\left\|\hat{\boldsymbol{\theta}}_1 - \boldsymbol{\theta}^*\right\|^2\right)$ are approximated by averaging over $10^6$ independent sample squared errors. Results show that for $k = 1$, there is a significant advantage of SU over the Bernoulli. But as the sample size increases, this advantage fades out, as we expect given the theory of the asymptotic optimality of the Bernoulli distribution.

### 5. CONCLUSION

In this paper, we investigate the performance of a non-Bernoulli distribution (specifically, the segmented uniform) for perturbation vectors in one step of SPSA. We show that for certain choices of parameters, non-Bernoulli will be preferred to the Bernoulli as the perturbation distribution for one-iteration SPSA. Furthermore, results in numerical examples indicate that we may generalize the above conclusion to other small sample sizes too, i.e., to two or more iterations of SPSA. In all, this paper gives a theoretical foundation for choosing an effective perturbation distribution when $k = 1$, and numerical experience indicates favorable results for a limited range of values of $k > 1$. This will be useful for SPSA-based optimization process for which available sample sizes are necessarily small in number.